\documentclass[%
 superscriptaddress,
 reprint,
 showpacs,preprintnumbers,
 nofootinbib,
 amsmath,amssymb,
 aps,
 prd,
 floatfix,
]{revtex4-1}

\usepackage{graphicx}

\usepackage{xcolor}

\usepackage{lipsum}

\usepackage{amssymb,amsmath,amstext,amsthm,amsfonts,amsfonts}
\usepackage[utf8]{inputenc}
\usepackage{subcaption}
\usepackage[linesnumbered,ruled]{algorithm2e}
\usepackage{url}
\usepackage{soul}
\usepackage{bm}
\usepackage[normalem]{ulem}
\usepackage{multirow, makecell}

\DeclareMathAlphabet{\mathsfit}{\encodingdefault}{\sfdefault}{m}{sl}
\SetMathAlphabet{\mathsfit}{bold}{\encodingdefault}{\sfdefault}{bx}{n}

\begin{document}
	\title{Uncovering the neutrino mass ordering with the next galactic core-collapse supernova neutrino burst using water Cherenkov detectors}

    \author{C\'esar Jes\'us-Valls}
    \email[E-mail: ]{cesar.jesus-valls@ipmu.jp}
	\affiliation{Kavli IPMU (WPI), UTIAS, The University of Tokyo, Kashiwa, Chiba 277-8583, Japan}
    
\begin{abstract}
\noindent
A major challenge of particle physics is determining the neutrino mass ordering (MO). Due to matter effects, the flavor content of the neutrino flux from a Core-Collapse Supernova (CCSN) depends on the true neutrino MO resulting in markedly different energy and angle distributions for the measured lepton in water Cherenkov neutrino detectors. In this article, those distributions are compared for eight different CCSN models and used to study how their differences affect  the determination of the neutrino mass ordering. In all cases, the inferred neutrino mass ordering is found to be either correct or inconclusive, with no significant false positives. However, the substantial variation observed among model predictions emphasizes the criticality of ongoing research in CCSN modeling.

\end{abstract}

\maketitle

\section{Introduction}

\label{sec:introduction}
Neutrinos are essential ingredients of particle physics, astrophysics and cosmology, and yet, some of their fundamental properties remain unknown. In the three-flavor picture describing experimentally observed neutrino mixing, there are seven parameters divided into three neutrino masses ($m_1$,$m_2$ and $m_3$) and four mixing parameters ($\theta_{12}$, $\theta_{23}$, $\theta_{13}$ and $\delta_{CP}$). Currently, $\theta_{12}$, $\theta_{23}$, $\theta_{13}$, $\delta m^2_{21} (\equiv m^2_2-m^2_1$) and $|\Delta m^2_{32}| (\equiv |m^2_3-m^2_2|$) are known with good precision~\cite{deSalas:2020pgw,Esteban:2020cvm} so that only three quantities remain elusive: a) the absolute neutrino mass scale, for a review see Ref.~\cite{Formaggio:2021nfz}, b) the value of $\delta_{CP}$~\cite{Esteban:2020cvm,T2K:2019bcf} and c) the sign of $\Delta m^2_{32}$, namely, the so-called neutrino mass ordering (MO) that can be normal $m_2<m_3$ (NMO) or inverted $m_3 < m_2$ (IMO).\\
Regarding the MO, current data shows a statistical preference for the NMO of about 3$\sigma$~\cite{Esteban:2020cvm}. Future experiments Hyper-Kamiokande~\cite{Hyper-Kamiokande:2021frf}, JUNO~\cite{JUNO:2015zny} and DUNE~\cite{DUNE:2020ypp} will characterize the MO in detail via the study of atmospheric, reactor and accelerator neutrino oscillations. The complementarity of the above experiments is an attractive advantage: if all the MO results are consistent the existing view of neutrino phenomenology will be reinforced, however, the appearance of tensions in the data could give us hints of new physics~\cite{Denton:2020uda,Chatterjee:2020kkm, Chang:2022aas}, requiring theoretical extensions. In the second scenario, the presence of additional data would be particularly helpful in disentangling the true MO of new physical effects. In this regard, a unique opportunity would arise in the event of a galactic supernova (SN) neutrino burst ~\cite{Dighe:1999bi, Scholberg:2017czd, Brdar:2022vfr}.

\subsection{Supernova neutrino bursts}
Core collapse Supernovae (CCSNe) emit $O(10^{53})$~erg as neutrinos, O($10^{58}$) neutrinos of $\langle E_\nu\rangle \sim 10$~MeV, in a ten-second burst~\cite{Barwick:2004ep, Janka:2012wk}. A low galactic rate of $1.63 \pm 0.46$ CCSNe per century is expected~\cite{Rozwadowska:2020nab}. Consequently, so far only a couple dozen neutrinos have been detected from a single supernova, SN-1987a ~\cite{Kamiokande-II:1987idp,Hirata:1988ad,Bionta:1987qt,Bethe:1990mw}. Nonetheless, the detection of SN-1987a confirmed our basic understanding of CCSNe explosions and signified the start of experimental neutrino astrophysics. Furthermore, since the predicted neutrino flux is greatly influenced by a plethora of effects, the SN-1987a data was used to establish significant limits on several exotic processes~\cite{Raffelt:1987yt,Turner:1987by,Raffelt:2006cw,Chang:2016ntp, Chang:2018rso} and neutrino properties, including their mass~\cite{Bahcall:1987nx,Spergel:1987ex}, magnetic moment~\cite{Barbieri:1988nh} and flavor mixing~\cite{Jegerlehner:1996kx}. These achievements, however, are just a tantalizing hint of what might be possible with the next generation of neutrino detectors. The increase of available data would be spectacular, e.g., $O(10^{4-6})$ detected neutrinos at Hyper-Kamiokande for a burst at a distance of 10-1~kpc~\cite{Hyper-Kamiokande:2021frf}. Such drastic improvement would certainly pose new challenges, specifically, the decrease in statistical error would force a shift of the analysis focus towards the treatment and evaluation of systematic uncertainties and potential model biases, so far mostly overlooked, in order to get the most out of the precious supernova data.\\

\subsection{Experimental prospects: DUNE, Super-Kamiokande and Hyper-Kamiokande}
Among future experiments, DUNE is mainly sensitive to the $\nu_e$ flavor, such that the (non-)observation of the so-called neutronization peak in the first milliseconds of the explosion would naturally constitute strong evidence of the IMO (NMO)~\cite{Scholberg:2017czd}. Notably, the emission in this initial stage of the explosion is also the better understood from a phenomenological standpoint and therefore this analysis strategy can be considered robust~\cite{Scholberg:2017czd,DUNE:2020zfm}. In contrast, water Cherenkov detectors are mainly sensitive to the $\bar{\nu}_e$ flavor and, consequently, the positive identification of the neutronization peak would need to be accompanied of a more intricate analysis of the measured lepton kinematic distributions, as highlighted in the Hyper-Kamiokande design report~\cite{Hyper-Kamiokande:2018ofw}. In this article, the physics potential of such an analysis is discussed.

\subsection{Neutrino flux predictions}
The explosion mechanism of CCSN is still poorly understood~\cite{Burrows:2020qrp}. However, the gradual increase in available computational power has allowed the set of simplifying assumptions about said mechanism to be reduced over time, and in recent years CCSN models have begun to achieve realistic self-triggered explosions~\cite{Burrows:2020qrp}. Furthermore, a general concordance has emerged among the neutrino flux predictions from different research teams~\cite{Lentz:2015nxa,Melson:2015tia,Skinner:2018iti,OConnor:2018tuw,Kuroda:2020bdq}. New data from a future CCSN would be game-changing to better understand the dynamics of the explosion. Hyper-Kamiokande will have great sensitivity to discriminate between different explosion models~\cite{Hyper-Kamiokande:2021frf,Olsen:2022pkn} and might be complemented by studies from JUNO~\cite{Birkenfeld:2020etm} and DUNE~\cite{DUNE:2020zfm}.


\subsection{Flavor transformations}
\label{sec:fl_trans}
During the neutronization burst or shock period, i.e. the first $\lesssim 50$~ms~\cite{Scholberg:2017czd}, the matter potential is anticipated to be dominant over the neutrino-neutrino potential. Consequently, flavor transformations can be described by the standard Mikheyev-Smirnov-Wolfenstein (MSW) effect~\cite{Mikheyev:1985zog,Wolfenstein:1977ue, Scholberg:2017czd, DUNE:2020zfm, Hyper-Kamiokande:2021frf}. Since $\sin^2\theta_{13}>10^{-3}$~\cite{ParticleDataGroup:2020ssz}, a non-oscillatory adiabatic flavor conversion is expected~\cite{Smirnov:2004zv}, sensitive to the neutrino MO. At later times, $50 \lesssim t \lesssim 200$~ms~\cite{Scholberg:2017czd}, during the so-called accretion phase and along the cooling phases, that describe the remainder of the burst, non-trivial effects such as SASI (standing accretion shock instability), turbulence, and neutrino self-interactions might change significantly the flavor composition of the flux~\cite{Sawyer:2005jk,Nagakura:2021hyb,Padilla-Gay:2021haz,nagakura2021core,Burrows:2020qrp}.

\subsection{Relevant interaction cross-sections}
\label{sec:xsec_ch}
Supernova neutrino energies are typically  of the order of a few tens of MeV~\cite{Bethe:1990mw}. At such energies, the main interactions with detector targets consist of neutrino- and antineutrino-electron elastic scattering (eES)~\cite{Bahcall:1995mm}, electron antineutrino inverse beta decay (IBD)~\cite{Strumia:2003zx} with unbound protons such as Hydrogen in water, and neutrino-nucleon charged- and neutral-current interactions with bound nucleons (e.g. $\nu_e$-CC $^{16}$O~\cite{Suzuki:2018aey,Nakazato:2018xkv}). Due to nuclear effects, neutrino interactions with bound nucleons are poorly understood~\cite{VanDessel:2019obk,Gardiner:2020ulp}, instead, eES and IBD predictions are well known.

\section{Methodology}

\subsection{Flux models}
\label{sec:flux_models}
\begin{figure}[ht!]
    \centering
    \includegraphics[width=0.99\linewidth]{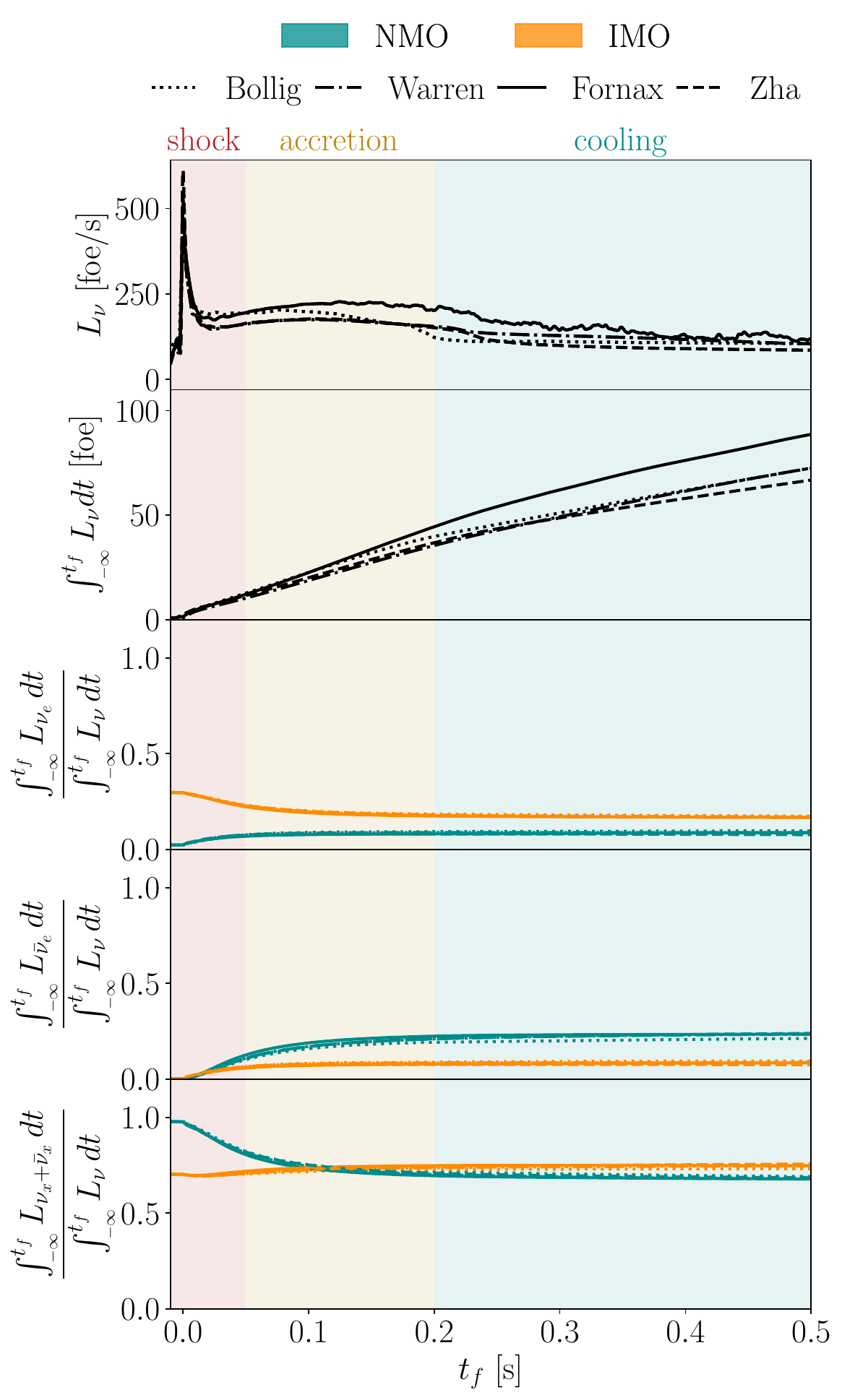}
    \caption{From top to bottom: Neutrino luminosity, cumulative total luminosity, cumulative fraction of the $\nu_e$ flavor, cumulative fraction of the $\nu_{\bar{e}}$ flavor and cumulative fraction of the $\nu_{x}+\nu_{\bar{x}}$ flavor as functions time. The cumulative luminosities are integrated from the start of the SN explosion, denoted by $-\infty$, until a time, $t$, for four different flux models. The different phases of the explosion are denoted by color shades. To simplify the plot visualization only a subset of all the flux models in the analysis is presented.} 
    \label{fig:flux_aligned}
\end{figure}

To study different flux models, the open-sourced software package \texttt{SNEWPY}~\cite{SNEWS:2021ewj,Baxter:2021xyq} is used. The models under consideration are, following \texttt{SNEWPY}'s nomenclature, \texttt{Bollig 2016}~($27~M_\odot$)~\cite{Mirizzi:2015eza},  \texttt{Fornax 2021}~($20~M_\odot$)~\cite{Burrows:2020qrp}, \texttt{Kuroda 2020}~($9.6~M_\odot$)~\cite{Kuroda:2020pta}, \texttt{Nakazato 2013}~($20~M_\odot$)~\cite{Nakazato:2012qf}, \texttt{OConnor 2015}~($40~M_\odot$)~\cite{OConnor:2014sgn},  \texttt{Sukhbold 2015}~($9.6~M_\odot$)~\cite{Sukhbold:2015wba}, 
\texttt{Warren 2020}~($13~M_\odot$)~\cite{Warren:2019lgb} and 
\texttt{Zha 2021}~($16~M_\odot$)~\cite{Zha:2021fbi}. This set of models aims to reflect the variability in the existing neutrino predictions among different models, computational approaches and progenitor masses. To account for flavor transformations, the neutrino flux predictions are modified according to \texttt{AdiabaticMSW} transformations\footnote{Implementation details are available in Appendix A of Ref.~\cite{SNEWS:2021ewj}.} using \texttt{SNEWPY}. These transformations depend on $\theta_{13}$ and $\theta_{23}$~\cite{Dighe:1999bi}, with values chosen from the Particle Data Group~\cite{ParticleDataGroup:2018ovx}. Since the uncertainty on these parameters is small the variations inflicted to the expected flavor predictions are minor, especially when compared to CCSN model-to-model variations. Henceforth, the uncertainty on these parameters is neglected, such as in Ref.~\cite{Hyper-Kamiokande:2021frf}.
\\Neutrino flux models, as seen in Fig.~\ref{fig:flux_aligned}, have the neutrino luminosity divided into four flavor categories: $L_{\nu_e}$, $L_{\bar{\nu}_e}$, $L_{\nu_x}$ and $L_{\bar{\nu}_x}$, where $x\equiv\mu+\tau$. The time evolution of the supernova explosion is markedly different between models such that $L(t)$ is not a model-robust observable. This is also true for the total neutrino luminosity integrated over time ($\int L_{\nu}(t)\,dt$) due to, among other effects, scale differences, such as the progenitor mass. However, as presented in Fig.~\ref{fig:flux_aligned}, the time-integrated fraction for each flavor is consistently different as a function of the true neutrino MO across flux models.
\\Since each model uses a different time reference definition, small time offsets are applied by setting the $t_0$ of each model as the time at which the neutrino luminosity reaches its maximum. Notably, this time frame could also be set for data by analyzing the time spectrum of the events. Although this calculation would contribute to the detector systematic uncertainty its role is later neglected as it is arguably a sub-leading correction\footnote{If a low number of interactions is recorded, statistical errors dominate. Else, $\sigma_{t_0}$ could be determined precisely, and a small time offset would translate into a minor variation of the integrated flavor composition, due to the small flavor gradients in the cumulative distributions observed at around 50~ms in Fig.~\ref{fig:flux_aligned}.}.

\subsection{Observable definition}
\begin{figure}[ht!]
    \centering
    \includegraphics[width=0.99\linewidth]{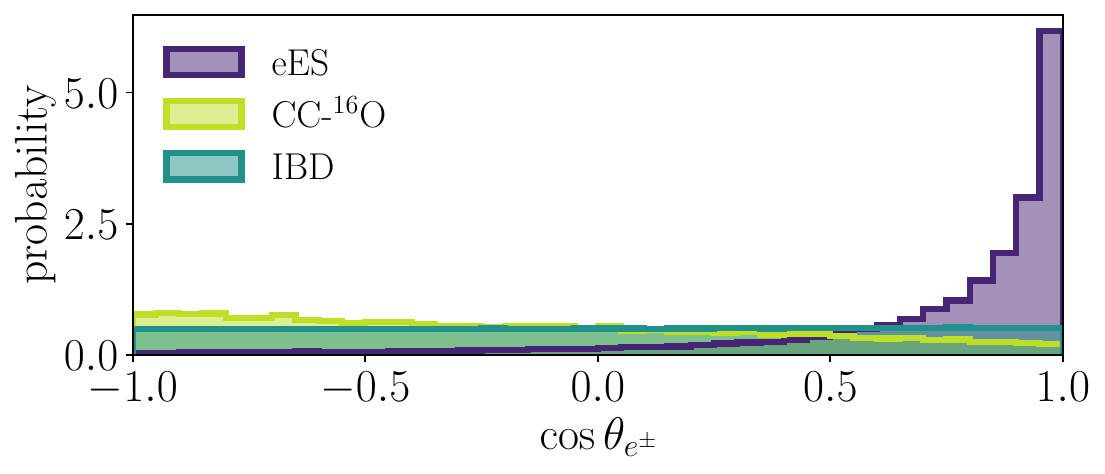}
    \caption{Binned probability distribution of the reconstructed angle for outgoing electrons and positrons,  split by interaction channel, generated for neutrino interactions in water using the 
    \texttt{Warren} flux model, presented in the text, transformed by \texttt{AdiabaticMSW} for the NMO.}
    \label{fig:costheta_spectrum}
\end{figure}
To be sensitive to the MO, it is necessary to define an observable that changes with the flux flavor content such that the differences presented in the former section can be quantified. From among all the interaction channels listed in Sec.~\ref{sec:xsec_ch}, all flavors contribute to eES whereas IBD is only possible for $\bar{\nu}_e$. Therefore, a measurement able to discriminate one reaction from the other is directly informative of the flux flavor and therefore of the true neutrino MO. It is well known that electrons emitted in eES are peaked forward whereas those from IBD and CC-$^{16}$O are quite isotropic, as illustrated in Fig.~\ref{fig:costheta_spectrum}. This feature has been used for years in the study of solar neutrinos~\cite{Super-Kamiokande:2016yck} and provides the best-known strategy to reconstruct the CCSN position in the sky~\cite{Hyper-Kamiokande:2018ofw}. Here, this feature is used to study the neutrino MO by noting that the expected shape of the angular distribution of outgoing $e^{\pm}$ is strongly influenced by the flux flavor content and therefore by the true neutrino MO. In addition, most interactions are due to eES at low energies and its proportion decreases in favor of IBD for increasing energies, as it can be seen in Fig.~\ref{fig:e_reco_cut} discussed later. Thus, binning the outgoing-lepton kinematic distributions in two dimensions (angle and energy), provides additional information on the neutrino flux flavor composition and increases the sensitivity to the neutrino MO. Time information, which varies considerably with the explosion model, is only used to select events detected within the first $50$~ms, corresponding to the neutronization phase. To minimize the influence of the flux normalization, earlier presented in Fig.\ref{fig:flux_aligned},  the energy and angle distribution is normalized to unity, making the analysis to rely exclusively on the distribution shape.

\subsection{Super-Kamionade and Hyper-Kamiokande}
Super-Kamiokande, 25~kT of fiducial mass, is currently the world reference water Cherenkov neutrino detector. Hyper-Kamiokande, 187~kT of fiducial mass, is being built nearby and it is expected to take over in 2027. The reconstruction performance of Super-Kamiokande is very well understood as Super-Kamiokande has been operating for nearly three decades. The performances of Hyper-Kamiokande are expected to match or surpass those of Super-Kamiokande~\cite{Hyper-Kamiokande:2018ofw}. Therefore, to simulate detector effects in both Super-Kamiokande and Hyper-Kamiokande the true lepton energy and angle predicted by \texttt{sntools} are smeared into a collection of reconstructed energies and angles using the reconstruction performances reported by the solar neutrino analysis of Super-Kamiokande~IV~\cite{Super-Kamiokande:2016yck} and used as later described in Sec.\ref{sec:eva_syst_unc} to account for the detector uncertainty. Other detector effects that might play a role when the data collection is time-sparse,  e.g. coincidental backgrounds and the trigger efficiency, can be safely neglected for a CCSN given that all events are collected in a few seconds.

\subsection{Event simulation and selection criteria}
\label{sec:methodology}
To translate neutrino luminosities into event rates the  \texttt{sntools}~\cite{Migenda:2021hnl} open-source software package is used. \texttt{sntools} reads fluxes in \texttt{SNEWPY} format and combines them with neutrino cross-sections to generate realistic event rates and particle distributions using Monte Carlo sampling.
\begin{figure}[ht!]
    \centering
    \includegraphics[width=0.99\linewidth]{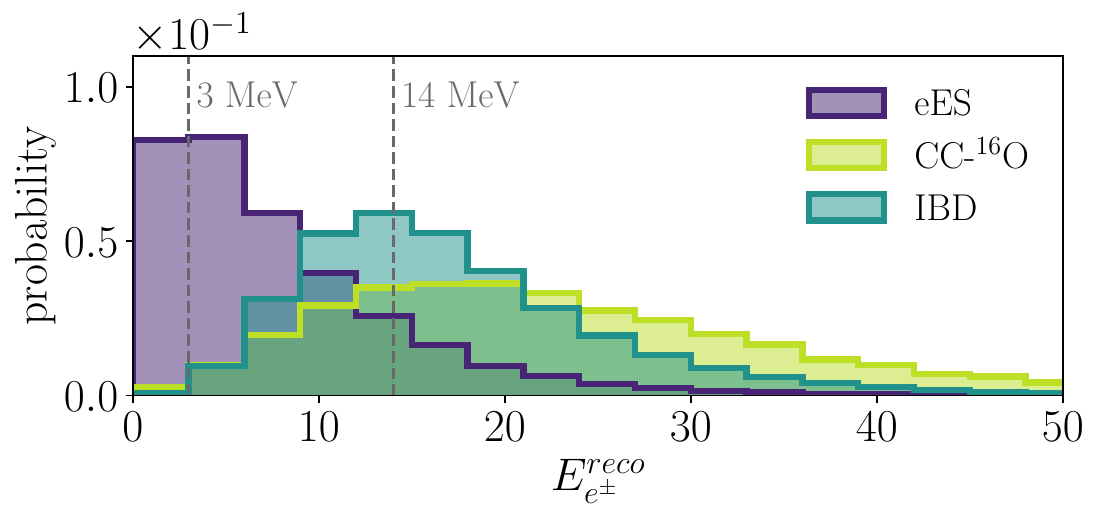}
    \caption{Binned probability distribution of the reconstructed energy for outgoing electrons and positrons, split by interaction channel, generated for neutrino interactions in water using the \texttt{Warren} flux model, presented in the text, transformed by \texttt{AdiabaticMSW} for the NMO.}
    \label{fig:e_reco_cut}
\end{figure}
\begin{table}[htb]
\footnotesize
\begin{tabular}{|c|ccc|ccc|}
 \hline
 \hline
  & \multicolumn{3}{c|}{NMO}    & \multicolumn{3}{c|}{IMO}  \\ 
  Cuts & IBD [\%] & eES [\%] & $^{16}$O [\%] &  IBD [\%] & eES [\%] & $^{16}$O [\%] \\
 \hline
A & 87.0$\pm$1.1  & 9.6$\pm$0.9 & 3.4$\pm$1.7 & 86.0$\pm$1.1 & 10.6$\pm$0.9 & 3.4$\pm$1.7  \\
B & 89.2$\pm$1.3 &  7.4$\pm$0.7 & 3.4$\pm$1.7 & 88.3$\pm$1.3 & 8.2$\pm$0.6 & 3.5$\pm$1.7 \\
C & 83.2$\pm$2.0 & 15.1$\pm$1.6 & 1.6$\pm$0.6  & 72.7$\pm$3.0 & 24.8$\pm$2.2 & 2.5$\pm$1.0  \\
 \hline
 \hline
 
\end{tabular}
\caption{Fractions of selected interactions in HK after applying the cuts \textit{A,B} and \textit{C}, corresponding to: \textit{A}) No cuts; \textit{B}) $E^{reco}_{e^\pm}>3$~MeV and \textit{C})  $14>E^{reco}_{e^\pm}>3$~MeV. The intervals reflect the mean and standard deviation of the values across all models with the exception of the Nakazato model\footnote{Pre-cut lepton energy predictions for the Nakazato model ($\langle E\rangle=35.4$~MeV) are completely different to those in all other models ($\langle E\rangle=17.3\pm1.6$~MeV, here the error describes the spread of the mean across the models) resulting in very different interaction fractions.}.}
\label{tab:cuts_fracts}
\end{table}
\\Over a million events are simulated for each SN model and MO and the expected normalized angular distribution of the outgoing $e^{\pm}$ is computed relative to the SN direction. The true $\theta_{e^\pm}$ and $E_{e^\pm}$ are smeared according to Super-Kamiokande~IV performances~\cite{Super-Kamiokande:2016yck}. Then, events with  $14>E^{reco}_{e^\pm}>3$~MeV are selected. This choice aims to: 1) account for a realistic detection energy threshold\footnote{In SK-IV a threshold as low as 3.49~MeV was used to study solar neutrinos~\cite{Super-Kamiokande:2016yck}. Hyper-Kamiokande expects to detect 10 photo-electrons/MeV compared to the 6 photo-electrons/MeV at SK-IV~\cite{Hyper-Kamiokande:2022smq}. Thus, a 3~MeV threshold is assumed.} and 2) increase (decrease) the fraction of eES (CC-$^{16}$O), which is concentrated at low (high) energies, see Fig.~\ref{fig:e_reco_cut}, and reduce the flux model-to-model discrepancies, particularly relevant at high $E^{reco}_{e^\pm}$.  The interaction fractions before and after the cuts are summarized in Tab.~\ref{tab:cuts_fracts}. As intended, the high energy cut significantly accentuates the difference in the predictions of eES events for each mass ordering.

\subsection{Evaluation of systematic uncertainties}
\label{sec:eva_syst_unc}
For each flux model and true neutrino MO the simulated distributions of reconstructed energy and angle are binned. For each event the true charged lepton energy and angle are smeared according to the SK-IV detector performances, and the prediction variation on each bin is taken as a detector systematic uncertainty. Similarly, to account for the cross-section model uncertainty in every bin, the weight associated to CC-$^{16}$O is varied assuming a 100\% normalization uncertainty\footnote{The absence of experimental data prevents the use of a more specific choice. The value of 100\% is a usual conservative choice when a normalization constraint is missing.}.
\begin{figure*}[ht!]
    \centering
    \includegraphics[width=0.32\linewidth]{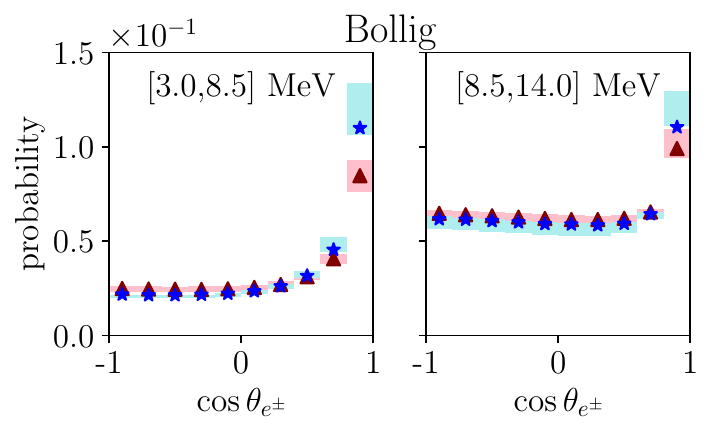}
    \includegraphics[width=0.32\linewidth]{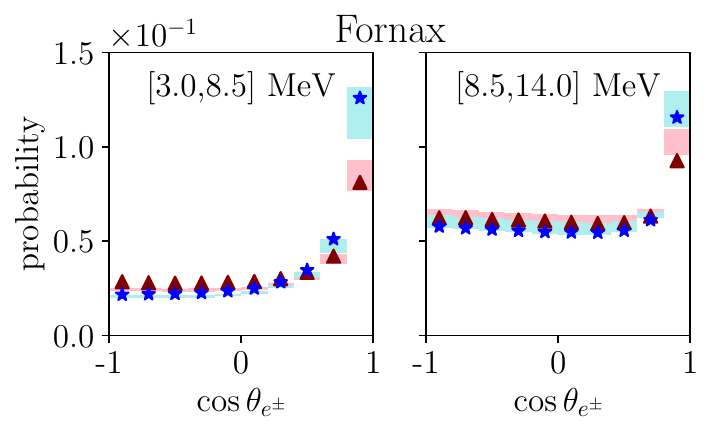}
    \includegraphics[width=0.32\linewidth]{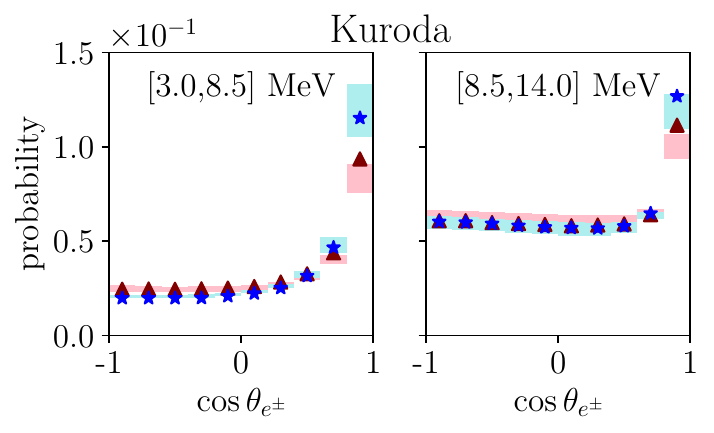}

    \includegraphics[width=0.32\linewidth]{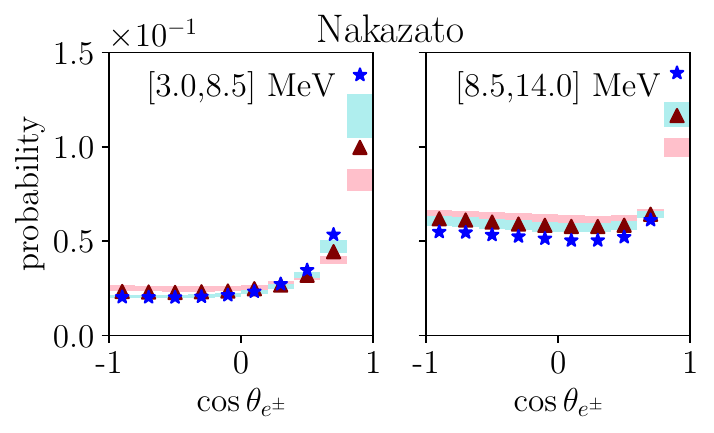}
    \includegraphics[width=0.32\linewidth]{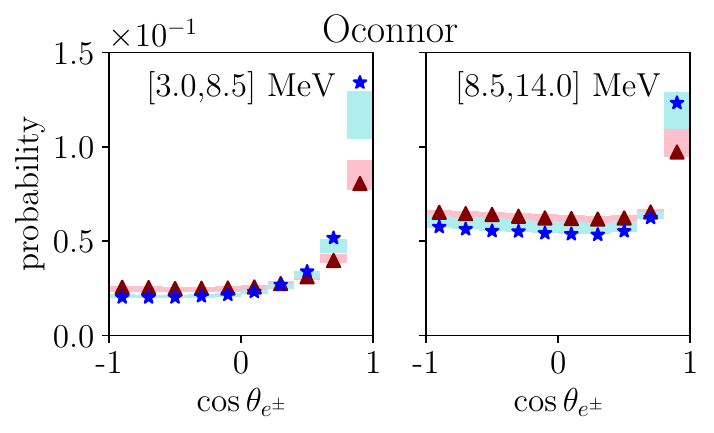}
    \includegraphics[width=0.32\linewidth]{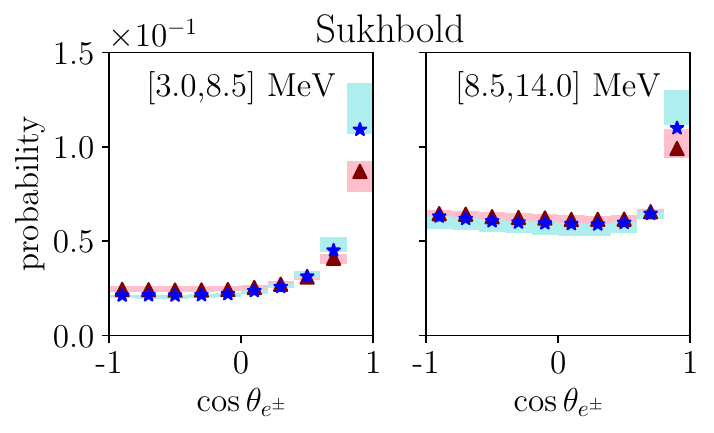}
    
    \includegraphics[width=0.32\linewidth]{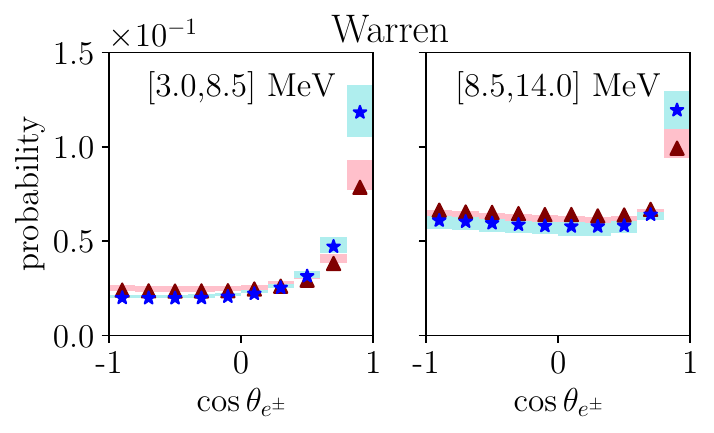}
    \includegraphics[width=0.32\linewidth]{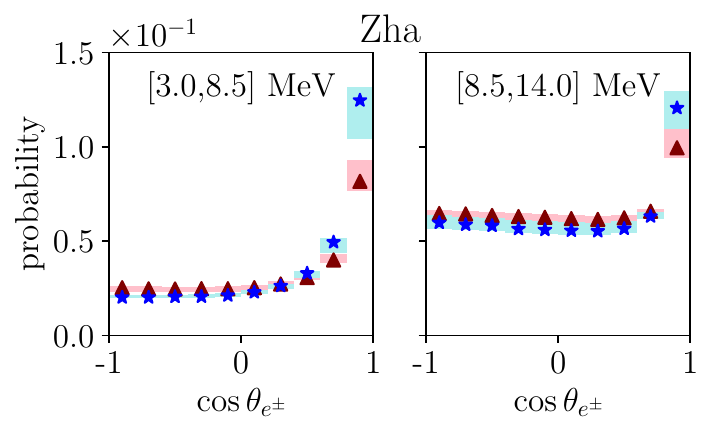}

    \caption{Individual model predictions for the reconstructed charged lepton energy and angle binned probability distributions are compared to the averaged predictions of all other models. Triangles (stars) indicate the predictions for the model in the title for the NMO (IMO). The binned predictions for all other models are represented by boxes. Each box central value (height) corresponds to the average (standard deviation) of the other models' prediction for the bin. The box height also includes the spreading introduced by the detector and cross-section uncertainties. Red (blue) boxes correspond to the NMO (IMO).} 
    \label{fig:model_patterns}
\end{figure*}

\subsection{Discriminating the MO from simulated data}
The discrimination confidence is calculated as a function of the number of selected neutrinos. To calculate it, the following method is used:
\begin{itemize}
    \item The binning scheme is defined. In total 20 bins are used: 2 energy bins, [3,8.5]~MeV and [8.5,14]~MeV, and 10 equally spaced angular bins from $-1 < \cos\theta_{e^\pm} < 1$. The choice is made to have a sufficiently large number of events per bin to apply Wilks' theorem~\cite{Wilks:1938dza}.
    \item The prediction of a CCSN model, $A$, for a given MO is treated as data and compared to both MO predictions for another model, $B$.
    \item For a given number of selected neutrinos  $N^{sel}$, the value of $\chi^2_{_{NMO}}$ and $\chi^2_{_{IMO}}$ is calculated using
    \begin{align}
        \chi^2_{_{M}}=\sum^{bins}_i \frac{(\mu^i_A-\mu^i_{B,M})^2}{\sigma_{_{M,B}}^2},\quad
        \chi^2_{_{M}}=\{\chi^2_{_{NMO}}, \chi^2_{_{IMO}}\}
    \end{align}
    where $A$ and $B$ denote the model choice and $M$ indicates the MO. The value of $\sigma^2$ consists of the addition in quadrature of the statistical uncertainty $\sigma_{stat}$, calculated from distributing $N^{sel}$ in each bin according to model $B$ predictions, and the combined systematic uncertainty of the detector and cross-section effects described in the previous section.
    \item Lastly, $\Delta \chi^2 \equiv \chi^2_{_{NMO}}-\chi^2_{_{IMO}}$ is calculated and translated into a number of $\sigma$, which corresponds to the discriminating variable referred to as $\Delta_{\textup{MO}}$.
   \end{itemize}

\begin{figure*}[ht!]
    \centering
    \includegraphics[width=0.99\linewidth]{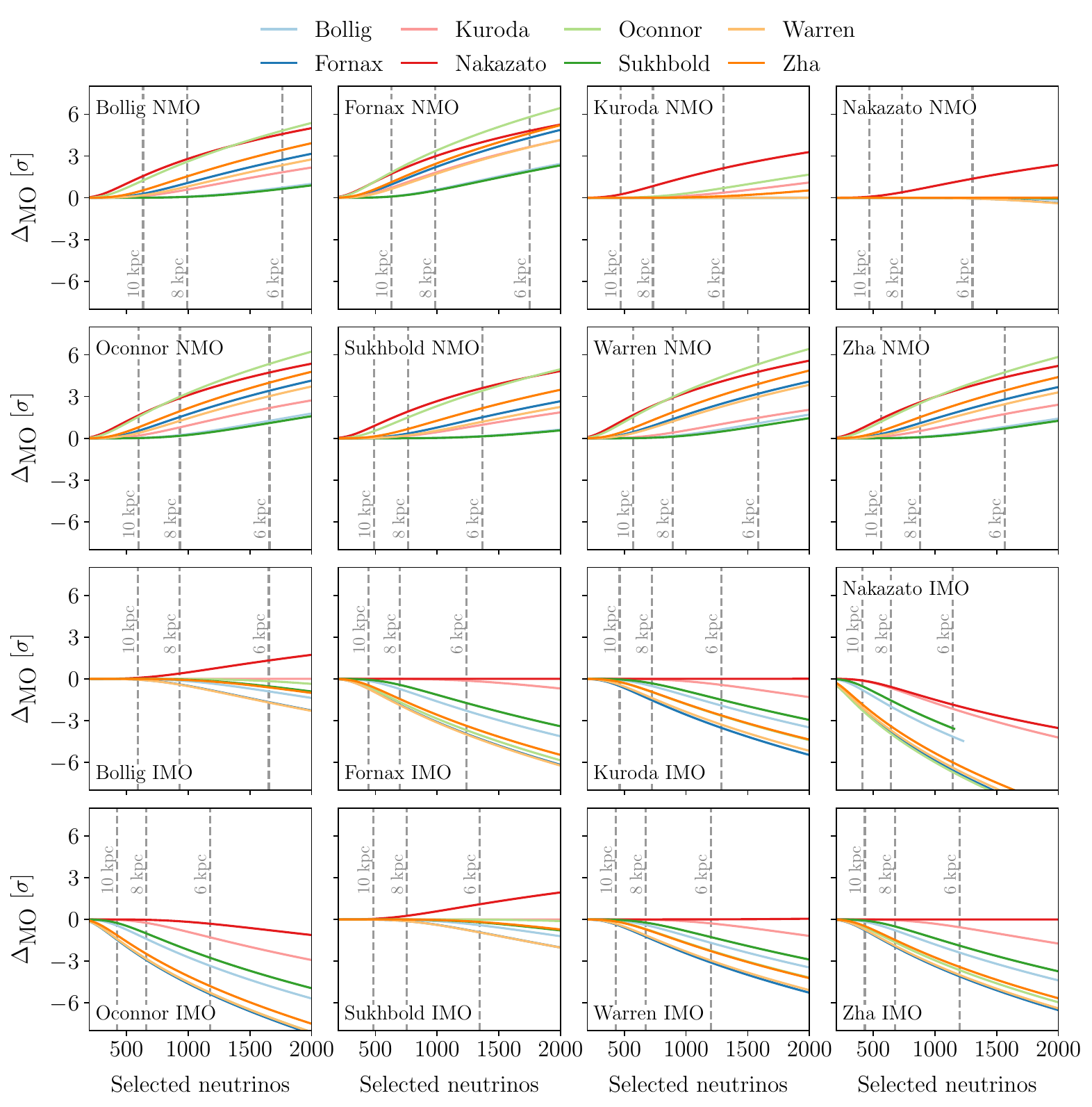}
    \caption{Neutrino mass ordering discrimination across all models. The model and MO written in every plot is the one for the simulated data. Solid lines indicate the mass ordering preference ($\Delta_{\textup{MO}}$) according to each model. Positive (negative) values of $\Delta_{\textup{MO}}$ correspond to the NMO (IMO). For a given model, $\Delta_{\textup{MO}}$ is only calculated as long as the predictions and the simulated data agree with a p-value~$>0.05$ for at least one of the predicted MOs. The x-axis indicates the number of selected neutrinos. To offer a visual reference, vertical dashed lines indicate the expected number of selected neutrinos in Hyper-Kamiokande for each model used as data at a CCSN distance of 10~kpc, 8~kpc and 6~kpc from Earth respectively.} 
    \label{fig:results_sensitivity}
\end{figure*}
\section{Results}
The 2D binned distribution of reconstructed energy and angle for the different CCSN flux models is presented in Fig.~\ref{fig:model_patterns}. For each model, the predicted distribution is compared to the averaged predictions of all other models as a function of the two neutrino mass orderings. Overall, the predictions for each MO are clearly distinct and follow a well-understood pattern. For the IMO, a significantly higher component of eES events is expected (Tab.~\ref{tab:cuts_fracts}) resulting in a larger proportion of forward going events when compared to the NMO predictions. For lower energies, the fraction of IBD events decreases (Fig.~\ref{fig:e_reco_cut}) resulting in a relatively smaller quasi-isotropic component. Overall, the models resemble the average of all other models, with the only exception being the Nakazato model, which noticeably departures from the common trend. This is not surprising, as discussed earlier in Tab.~\ref{tab:cuts_fracts}, the Nakazato model exhibits a completely different lepton energy distribution when compared to the rest of the models. In addition, the separation between the NMO and the IMO predictions is somewhat smaller for some models, e.g. Sukhbold, than for others, e.g. Oconnor.\\
Model predictions not agreeing reasonably well with data for any of the two MOs should not be used to draw conclusions about the neutrino MO as they could lead to biased results. Here, the simple approach of requiring a p-value~$>$0.05 for at least one of the two MOs is applied. However, in a real-case scenario, more sophisticated conditions could be explored to further reduce model dependencies.\\ 
Nevertheless, even when only using models showing plausible agreement, the CCSN data interpretation can be model dependent. This is because two models may agree relatively well in their predictions of one MO, but disagree significantly on the other. For instance, according to the Bollig model, the fraction of selected eES events is 15.2\% (21.8\%) for the NMO (IMO), while the Oconnor model predicts a proportion of 14.1\% (28.0\%) selected eES events for the NMO (IMO). To study the interplay among the different models, the confidence on the true MO is calculated for every model and mass ordering against all other CCSN models. The results are presented in Fig.~\ref{fig:results_sensitivity}. All models lead to either inconclusive results (Nakazato NMO, Bollig IMO and Sukhbold IMO) or align well with the true neutrino mass ordering. The results show that although estimating the statistical confidence of the MO in a model-independent way is currently challenging, the study of CCSN data in water Cherenkov detectors could provide plausible indications of the true neutrino MO. For most models, the indications become noticeable (strong) for a number of selected neutrinos approximately above 500 (1500), that translates, roughly, to a CCSN distance from Earth of 10~kpc (5~kpc) for Hyper-Kamiokande and about 3~kpc (1.5~kpc) for Super-Kamiokande.

\section{Conclusions}
A galactic CCSN neutrino burst would provide a once-in-a-generation opportunity to advance our knowledge on fundamental physics. In this article the potential of water Cherenkov detectors to infer the neutrino mass ordering from CCSN data has been studied using the flux predictions of eight different CCSN models. The results show that, even if the existing neutrino flux model-to-model discrepancies in the analyzed distributions are significant, the inferred neutrino mass ordering is in general well aligned with the simulated one across models. In the case of a galactic CCSN, Hyper-Kamiokande (Super-Kamiokande) is expected to be able to provide plausible indications of the true neutrino MO by analyzing the flavor composition of the neutrino flux for neutrino bursts at a distance of $\lesssim8$~kpc ($\lesssim 2.5$~kpc) from Earth.

\section*{Acknoledgments}
The author is grateful for reviewer comments that significantly enhanced the study. I acknowledge CERN for organizing the INSS where this project was sparked and I appreciate fruitful discussions with T.~Lux, E. Ramos-Casc\'on and P.~Barham as well as the feedback from C.~Alt, S.~Dolan, C.~Grimal-Bosch, S.~Julià-Farré and J.~Migenda.

\bibliographystyle{apsrev4-1}
\bibliography{biblio}

\end{document}